\documentclass{svproc}
\usepackage{marvosym}
\usepackage{amsmath}
\usepackage{graphicx}

\usepackage{url}
\usepackage{hyperref}
\usepackage{cite}
\usepackage{float}

\providecommand{\doi}[1]{doi:\discretionary{}{}{}#1}

\def\orcidID#1{\unskip$^{[#1]}$}
\def\letter{$^{\textrm{(\Letter)}}$}

\begin{document}
\mainmatter              % start of a contribution
\title{Superconductivity and trimers on attractive-$U$ Hubbard ladders}
\titlerunning{Superconductivity and trimers}  % abbreviated title (for running head)
\author{Ian Pil\'{e}\inst{1}\letter\orcidID{0000-0001-7805-7400} \and Evgeni Burovski\inst{1}\orcidID{0000-0001-8149-0483}}
\authorrunning{Ian Pil\'{e} and Evgeni Burovski} % abbreviated author list (for running head)
\tocauthor{Ian Pil\'{e} and Evgeni Burovski}
\institute{HSE University, Moscow, Russia \\
\email{pileyan@gmail.com, evgeny.burovskiy@gmail.com}
}

\maketitle              % typeset the title of the contribution
\begin{abstract}
We investigate the interplay between superconducting correlations and trimer formation in polarized two-component Fermi gases confined to multileg attractive-$U$ Hubbard ladders. Employing density matrix renormalization group (DMRG) simulations, we explore the effects of spin-dependent tunneling amplitudes on these systems. Specifically, we analyze how bound states of three fermions (trimers) impact Fulde-Ferrell-Larkin-Ovchinnikov (FFLO) superconducting correlations at commensurate charge carrier densities, where $2n_{\uparrow} = n_{\downarrow}$. In one-dimensional (1D) systems, trimer formation is known to suppress FFLO correlations exponentially. Our results demonstrate that this suppression persists on ladder lattices of small width, effectively mirroring the 1D behavior. However, we find a striking departure from the 1D regime as the ladder width increases. On ladders with a width of four legs, the influence of trimers on superconducting correlations becomes negligible, suggesting that wider ladder systems provide a distinct environment where FFLO-like pairing remains robust even in the presence of trimer states. These findings underscore the dimensional crossover in Hubbard systems and shed light on the mechanisms governing superconductivity and bound-state formation in strongly correlated fermionic systems. Our work has implications for understanding unconventional superconductivity in strongly correlated systems.

\keywords{Fulde-Ferrell-Larkin-Ovchinnikov state $\cdot$ Trimers $\cdot$ Hubbard ladders $\cdot$ DMRG}

\end{abstract}

\section*{Introduction}
 
Recent advances in the study of ultracold atomic gases open up new possibilities to solve fundamental theoretical problems in direct experiments \cite{Revelle2016}. Ultracold atomic systems, due to their high tunability and exceptional isolation from environmental noise, offer unique platforms for realizing and studying phenomena that are challenging to observe in condensed matter systems. These platforms enable the exploration of controlled environments where interactions, dimensionality, and other parameters can be precisely adjusted, making them ideal for probing fundamental physics. One of the unresolved questions in this domain concerns the existence of superconductivity in systems with unequal populations of fermions with opposite spin directions. Among the potential solutions, the Fulde-Ferrell-Larkin-Ovchinnikov (FFLO) state \cite{Fulde1964, Larkin1964} stands out as a promising candidate. In this exotic state, the superconducting order parameter becomes spatially modulated due to Cooper pair formation with non-zero center-of-mass momentum, providing a rich area for theoretical and experimental inquiry.

Experimental efforts to identify FFLO superconductivity in polarized atomic quantum gases have so far been limited to three-dimensional (3D) configurations \cite{Partridge2006}. These studies, while groundbreaking, have faced significant challenges, including the difficulty of achieving stable configurations under experimental conditions. A promising direction, however, involves confining atoms in elongated traps where the FFLO state is significantly more stable \cite{Orso2007}. This stability has been further corroborated by detailed numerical simulations for both strictly one-dimensional (1D) and quasi-one-dimensional geometries \cite{FHM2007_1D, PotapovaPile}. These studies underscore the importance of dimensionality and interaction strength in stabilizing FFLO phases, paving the way for future experimental realizations.

There is also a number of approaches aimed at providing indirect evidence \cite{Agterberg2020_cuprates_review} for FFLO-like physics in different experimental designs involving cuprates \cite{Du2020_cuprate1}, organic superconductors \cite{Imajo_organic2}, and pnictides \cite{Cho2017}. These efforts leverage the unique properties of these materials to identify signatures consistent with FFLO phases. Despite these advances, direct observation of the FFLO state remains elusive, primarily due to impurity scattering, which can completely suppress long-range ordering \cite{Song2019_multilayer_fflo}. The ultracold atomic gas platforms, however, circumvent many of these challenges, providing clean systems with controllable parameters to explore FFLO physics, thereby presenting an unparalleled opportunity to resolve longstanding theoretical questions.
Ultracold gases are particularly well-suited for experimental modeling of the Hubbard model. These experimental designs can mimic the Hubbard model on different lattice geometries with a broad range of tunable parameters, owing to unprecedented control over lattice geometry, interaction strengths, and tunneling amplitudes \cite{Esslinger}. Various geometries have been explored, including spherically symmetric 3D traps \cite{Zwierlein2006, Zwierlein2006_2}, elongated cigar-shaped traps \cite{Liao2010}, and arrays of cigar-shaped traps \cite{Revelle2016}. Additionally, recent progress has been made in creating optical analogues of Kagome \cite{Kagome} and Lieb \cite{Lieb1, Lieb2} lattices, which facilitate the study of unconventional physical states emerging near flat bands. Flat-band systems, in particular, exhibit unique quantum phenomena due to the high density of states and suppressed kinetic energy, making them a fertile ground for novel phases of matter. New theoretical developments \cite{IskinTheory, HeTheory} have also generalized the BCS-BEC (Bardeen-Cooper-Schrieffer - Bose-Einstein Condensate) crossover theory to two-band continuous models describing superfluid Fermi gases near the Feshbach resonance, significantly broadening the scope of accessible physics and offering insights into multiband effects in ultracold systems.

Another topic of considerable interest is the formation of bound states in Fermi gases with unequal masses, such as mixtures of \textsuperscript{6}Li and  \textsuperscript{40}K near the Feshbach resonance \cite{OrsoPitaevski, Huang}. These mass-imbalanced systems introduce additional degrees of freedom and complex interaction landscapes, which are crucial for understanding few-body and many-body physics. Experimental realizations of these systems include two-component Fermi gases in spin-dependent optical lattices, where effective masses are different \cite{Mandel}. These configurations provide opportunities to explore new regimes of few-body physics and bound state formation in mass-imbalanced systems, leading to the discovery of exotic bound states and effective interactions that are not present in mass-balanced counterparts.

The study of multimer formation, particularly in fermionic systems, has seen remarkable advancements in recent years. This field received a significant boost with the experimental realization of Efimov trimers consisting of three identical bosons in continuum \cite{Kraemer, Zaccanti, Hulet}. These findings laid the groundwork for investigating analogous phenomena in fermionic systems, where the interplay between quantum statistics and interaction symmetries gives rise to unique bound states. A notable advancement in fermionic trimer physics is the exploration of universal trimers in systems with p-wave interactions \cite{ChenGreene}. Researchers have studied the formation of trimers near p-wave unitarity and identified a novel phenomenon called the faux-Efimov effect, where van der Waals forces play a critical role. In this regime, trimer states emerge with symmetries distinct from those observed in s-wave interactions, expanding the range of possible quantum phases and their stability conditions.

Spin-orbit coupling introduces another avenue for stabilizing and manipulating trimer states. For instance, Qiu, Cui, and Yi \cite{QiuCuiYi} demonstrated the emergence of a universal trimer stabilized by the symmetry of the single-particle spectrum in a system comprising a spin-orbit-coupled Fermi sea interacting with an impurity. This study revealed that particle-hole fluctuations play a significant role in enhancing the stability of these trimers as the Fermi energy increases, providing insights into designing systems with robust trimer phases. The interplay of spin-orbit coupling and interaction anisotropy offers a versatile toolkit for exploring the dynamics and stability of complex few-body systems.

One-dimensional (1D) systems are particularly advantageous for studying spinless trimers due to the enhanced role of quantum correlations. Gotta et al. \cite{Gotta} explored trimer formation in a 1D chain of spinless fermions with correlated hopping terms. Their work identified the emergence of a two-fluid system, where a trimer fluid forms as a result of a gain in kinetic energy alongside a fluid of unbound fermions. These phases exhibited non-trivial stability properties, laying the groundwork for further studies on few-body physics in constrained geometries. Additionally, 1D systems provide an excellent platform for probing the interplay between quantum entanglement and bound state formation, which has implications for quantum information science and the development of novel quantum technologies.

The ground state of Bose-Hubbard systems with on- and off-site multimers has been extensively studied using both variational calculations and density matrix renormalization group (DMRG) simulations \cite{IskinKeles}. These studies highlight the intricate balance between interaction energy and tunneling processes in stabilizing multimers, providing valuable insights into the design of synthetic quantum materials. In other fermionic systems, the emergence of "trimeron" states immersed in a spin-polarized Fermi sea has been a subject of interest \cite{Ragheed}. It has been shown that in the 1D case, trimer formation suppresses superconducting correlations exponentially as long as the densities of charge carriers with opposite spins remain commensurate \cite{Orso2010trimers}. Earlier studies \cite{BurovskiOrsoJolicoeur, Mattis, Valiente} also explored the role of trimers in 1D systems, emphasizing their influence on superconducting and transport properties.
From a theoretical perspective, trimer formation has been extensively studied in systems with non-trivial lattice geometries. For example, on a sawtooth lattice near a flat band, researchers have identified unique mechanisms that stabilize trimer states \cite{OrsoSingh}. More recently, exact solutions to the trimer problem in generic multiband Hubbard models have been developed in the form of coupled integral equations \cite{IskinTri}. These findings provide a rigorous theoretical framework for exploring the role of trimer states in complex lattice systems, bridging the gap between few-body physics and many-body phenomena. Additionally, the insights gained from these studies are expected to have implications for understanding unconventional superconductivity and correlated phases in materials with non-trivial band topology.

This paper addresses the effects of trimers on superconducting correlations in the quasi-1D regime. The paper is organized as follows: first we define the model under consideration in Model and Methods section, then specify the parameters and the construction of initial state for DMRG in the Numerical simulations section and then proceed to results. We study the quasi-1D attractive Hubbard model with spin-dependent tunneling amplitude on wider ladders of up to four legs. To compute trimer's binding energy and correlation functions we use numerically unbiased density-matrix renormalization group (DMRG) simulations \cite{White1992, Schollwoeck2005}. Our numerical results for strict 1D is also consistent with previous simulations of Ref.\  \cite{Orso2010trimers}. By considering wider ladders, we find that the existence of a trimer gap is not only the feature of 1D-chain but also affects superconducting regime on 2-legged- and, arguably, 3-legged-ladders. Specifically, we find that (i) for spin independent tunneling amplitudes trimers are not formed on Hubbard ladders, and (ii) for largely different spin-dependent tunneling amplitudes the influence of trimers on superconductivity decreases with the transition from a strictly 1D regime to a quasi-1D. In addition, it is shown that with an increase in the occupation numbers, the trimer gap decreases, which indicates that the influence of trimers is present only at small occupation numbers.

%%%%%%%%%%%%%%%%%%%%%%%%%%%%%%%%%%%%%%%%%%%%%%%%%%%%%%%%%%%%%%%%%%%
\section*{Model and Methods}
We consider here the attractive-$U$ Hubbard model of a two-component Fermi gas at zero temperature on a $L \times W$ ladder defined by the Hamiltonian:

\begin{equation} \label{Hubbard}
\hat{H}=-t_{\sigma} \sum_{\langle ij \rangle, \sigma} \left( \hat{c}_{i,\sigma}^{\dagger} \hat{c}_{j, \sigma}^{\phantom \dagger} + {\rm h.c.} \right) + U \sum_{i} \hat{n}_{i, \uparrow} \hat{n}_{i, \downarrow} \ .
\end{equation}
$\hat{c}_{i,\sigma}^{\dagger}$($\hat{c}_{i,\sigma}$) are a creation (annihilation) operators of a fermion with pseudospin $\sigma=\uparrow, \downarrow$ on site $i$; $\hat{n}_{i, \sigma} = \hat{c}_{i,\sigma}^{\dagger} \hat{c}_{i,\sigma}^{\phantom \dagger}$ is the number-of-particles operator.
We set the hopping amplitudes $t_{\uparrow} = 1$ and $t_{\downarrow} \leq t_{\uparrow}$ varies (both are the energy scale) , and $U < 0$ is the local Hubbard attraction parameter between two fermions with opposite spins. The summation over $i$ in Eq.~\eqref{Hubbard} runs over $L \times W$ sites of the ladder lattice and we only take $W \ll L$ so that the model is quasi-1-dimensional.

For the model ~\eqref{Hubbard} ground state energies, $E_0(N_\uparrow, N_\downarrow)$ in the canonical ensemble are explicitly characterized by the numbers of spin-up and spin-down charge-carriers, $N_\uparrow$ and $N_\downarrow$, or equivalently by the filling fractions, $n_\sigma= N_\sigma / LW$. We can calculate the binding energy of a trimer for the system as follows:

\begin{equation} \label{eq:trimer}
E_{trimer} = E_0(1,2) - E_0(1,1) - E_0(0,1)
\end{equation}

The trimer we are considering is a bound state of one spin-up fermion and two spin-down fermions. 

The trimer binding energy was computed for various lengths and configurations of ladder lattices. The ground-state energy calculations were conducted for systems with a fixed number of electrons having spin-up and spin-down orientations. The key parameters varied across the analyzed systems included the ladder length, the type of transverse boundary conditions (either open or periodic), and the ratio of tunneling amplitudes for electrons with different spin directions. For multiple configurations, the relationship between the trimer binding energy and the lattice length was evaluated for \( L \) ranging from 50 to 250.

For $L \neq \infty$ one can expect some finite-size effects to appear, so we plot $E_{trimer}(1/L)$ and extrapolate this dependence to $1/L = 0$ using least-squares to find $E_{trimer}$ on a ladder of infinite size. These approximations are performed for both the balanced ($t_{\uparrow} = t_{\downarrow}$) and imbalanced ($t_{\uparrow} \neq t_{\downarrow}$) cases.

The other measure that can help us track trimer presence is the trimer gap, which is defined as follows:

\begin{equation} 
\label{eq:trimer_gap}
\begin{aligned}
\Delta_{trimer} = -E_0(N_{\uparrow}+1,N{\downarrow}+2) - E_0(N_{\uparrow},N_{\downarrow}) + \\
 E_0(N_{\uparrow}+1,N_{\downarrow}+1) + E_0(N_{\uparrow},N_{\downarrow}+1)
 \end{aligned}
\end{equation}

Previous results \cite{Orso2010trimers} for a single Hubbard chain suggest that the trimer gap should decrease as the filling fractions increase.

It was also shown \cite{PotapovaPile} that there are superconducting correlations of FFLO type in partially polarized phase on multileg attractive-U Hubbard ladders for balanced ($t_{\uparrow} = t_{\downarrow}$) case. We will denote superconducting correlation functions as:
\begin{equation} \label{eq:correlation}
\Gamma(\mathbf{x}, \mathbf{x}_0) = \langle \hat \Delta^\dagger(\mathbf{x}_0) \hat \Delta(\mathbf{x} + \mathbf{x}_0)\rangle
\end{equation}

Here $\hat \Delta(\mathbf{x}) = \hat c_{\mathbf{x}\uparrow} \hat c_{\mathbf{x}\downarrow}$ annihilates a pair of fermions at a lattice site with coordinates $\mathbf{x}$. In the 1D limit, $W=1$,  $\Gamma$ is expected \cite{Yang2001} to decay algebraically with $|\mathbf{x}|$ and oscillate with the typical FFLO momentum $Q = \pi (n_\uparrow - n_\downarrow)$. For ladder lattices the behaviour is qualitatively similar \cite{PotapovaPile}, but oscillation frequency is a bit different.

In active presence of trimers these correlations should decay exponentially \cite{Orso2010trimers} instead of algebraically, which shows that trimers suppress superconductivity.

\newpage 
%%%%%%%%%%%%%%%%%%%%%%%%%%%%%%%%%%%%%%%%
\section*{Numerical simulations}
To extract ground-state energies $E_0(N_\uparrow, N_\downarrow)$ and superconducting correlations $\Gamma(\mathbf{x}, \mathbf{x}_0)$ of the Hamiltonian \eqref{Hubbard} for the states with fixed number of electrons we use DMRG computations as implemented in iTensor\cite{iTensor} package in C++ implementation. All tensor train operations necessary for DMRG simulations are available out of the box.  DMRG simulations of quasi-one-dimensional systems are notoriously computationally hard (complexity is exponential in transverse direction width), as the algorithm is one-dimensional by design, so energy and correlation values for wider ladders can only be obtained in considerably long time (days). Parallelization of Matrix-Product States-based algorithms heavily rely on the linear algebra library. We used OpenBLAS package for main runs. Parallelization is handled by OpenBLAS automatically and we use up to 32 threads. Computational experiments performed by running four hundred separate simulations. All simulations were performed on HSE HPC\cite{Kostenetskiy2021}. We use maximum bond dimension of and MPS up to $1800$ and keep energy convergence threshold around $10^{-7}$ for $W$ up to $3$ and $10^{-6}$ for $W=4,5$. 

\subsection*{Selection of Initial State}

Since the convergence speed of the DMRG algorithm can significantly depend on how close the initial conditions are to the specific final configuration of electrons, the following heuristic was used for the calculations. It is known that for a one-dimensional chain of identical charge carriers with different spin directions, the energy-minimizing state will be antiferromagnetic. For a lattice with specific $W,L$ and some fixed numbers of electrons $N_{\uparrow}$ $N_{\downarrow}$, the following initial state construction is applied:

\begin{itemize}
    \item We will enumerate the sites as shown in Fig. 1, first along $W$, and then along $L$.
    \item If $N_{\uparrow} + N_{\downarrow} \le WL$\\
    
    \begin{itemize}
        \item We will fill the lattice starting from site 1 in a checkerboard pattern until the electrons with the non-dominant spin direction are depleted.
        \item Then, the subsequent sites are filled with the remaining electrons with the dominant spin direction.
        \item If $N_{\uparrow} + N_{\downarrow} < WL$, then all sites starting from $N_{\uparrow} + N_{\downarrow}$ remain empty.
    \end{itemize}
    \item If $N_{\uparrow} + N_{\downarrow} > WL$
    \begin{itemize}
        \item Place the first $WL$ electrons following the algorithm of the previous section.
        \item Then, starting at site 1, add an additional electron with the opposite spin direction to the already occupied sites until all $N_{\uparrow} + N_{\downarrow}$ are used up, or until each site contains exactly 2 electrons.
    \end{itemize}
\end{itemize}
\vskip 10pt
\begin{figure}[hbt]
\centering
\includegraphics[width=0.7\columnwidth]{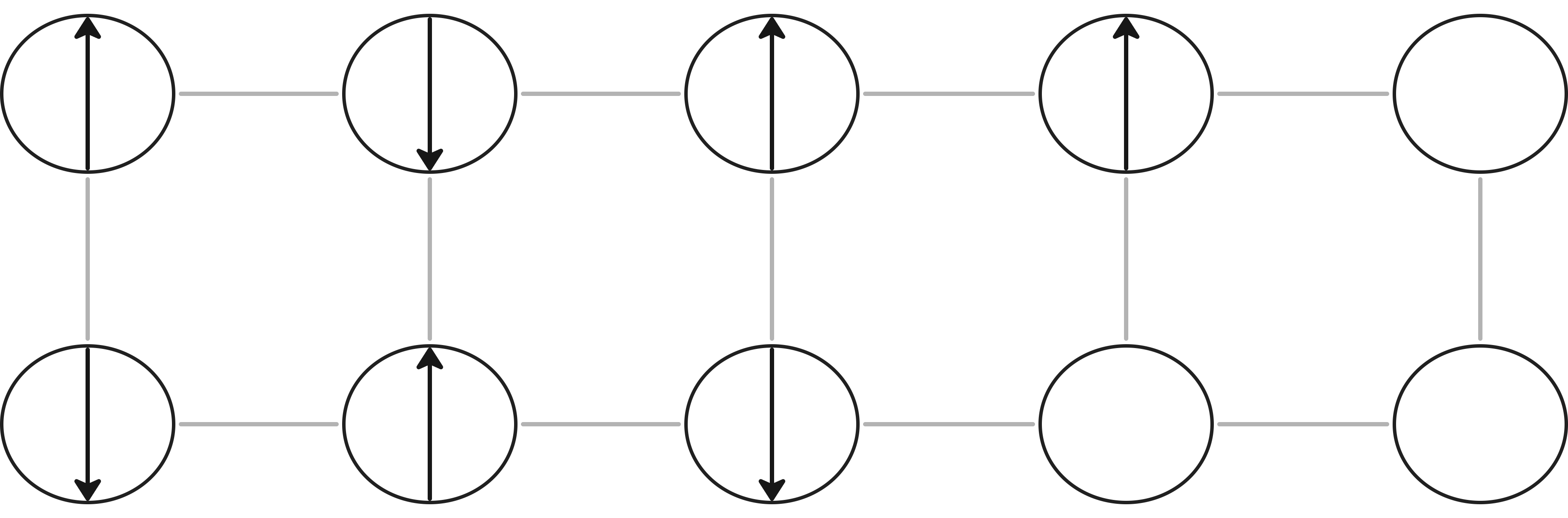}
\caption{Example of lattice filling with $W = 2$, $L = 5$, $N_{\uparrow} = 4$, $N_{\downarrow} = 3$ according to the algorithm described above.
\label{fig:checkerboard}
}

\end{figure}

%----------------------------------------------------------------------------------------
%	RESULTS 
%----------------------------------------------------------------------------------------

\section*{Results and discussion}
We first consider the balanced ($t_{\uparrow} = t_{\downarrow}$) case on ladder lattices of width $W= 3, 4, 5$ to probe that quasi-1D case does not differ much from strictly 1D chain (there are no trimers in 1D for balanced case). 

% width=0.99\columnwidth
\begin{figure}[H]
\includegraphics[width=0.49\textwidth, keepaspectratio=True]{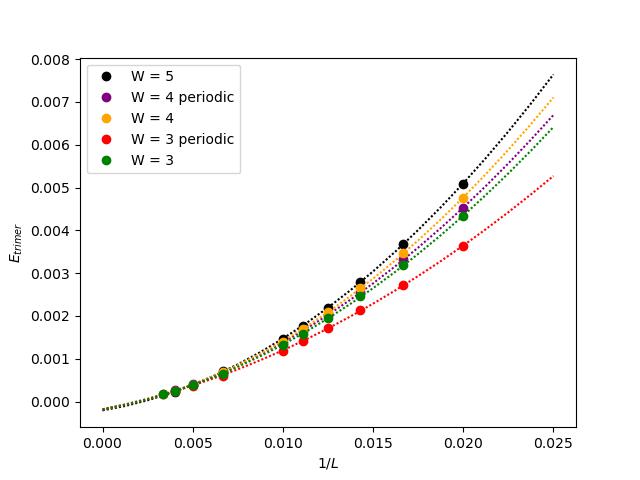}\quad 
\includegraphics[width=0.49\textwidth, keepaspectratio=True]{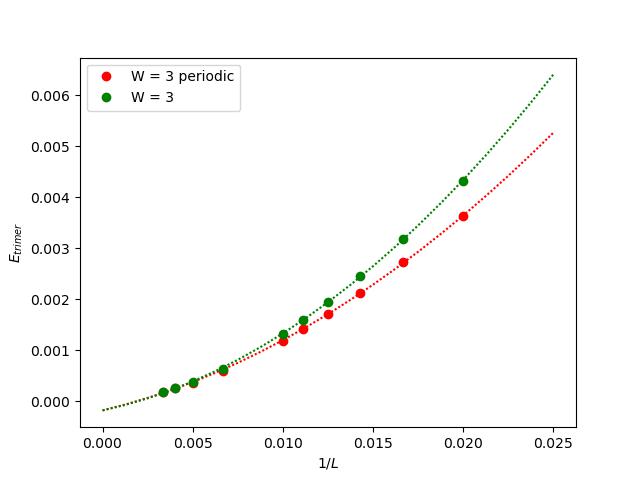} \\
\caption{\textbf{Left}: Binding energy of a trimer \eqref{eq:trimer} for $W = 3,4,5$, $U = -7$, and varying $L$. \textbf{Right}: Binding energy of a trimer \eqref{eq:trimer} for $W = 3$, $U = -7$, varying $L$ and open(green line) and periodic(red line) boundary conditions in transverse direction. All filled circles are DMRG results, and dotted lines are results of quadratic approximation to the limit $L \rightarrow \infty$ 
\label{fig:trimers_balanced}}
\end{figure}

Figure~\ref{fig:trimers_balanced} illustrates the relationship between \( E_{trimer} \) and \( 1/L \). The dots in the plot represent the results obtained from DMRG calculations, while the solid lines correspond to a quadratic approximation of the dependence as \( L \to \infty \). 
This approximation was implemented in Python using the least squares method provided by the Numpy\cite{numpy} library. By applying this method, we effectively extrapolate the behavior of the system to its infinite-length limit. 

When \( L \to \infty \), the computed values of \( E_{\text{trimer}} \) are on the order of \( 10^{-4} \), which is negligible and nearly indistinguishable from zero. This result indicates that any non-zero trimer energy values are artifacts arising from lattice corrections related to the finite size of the system. These corrections result from constraints imposed by the limited system length, causing minor perturbations in the trimer energy values. As a result, they remain slightly above zero but fall below the error margin of the coefficient approximation.  
It is noteworthy that the scale of these finite-length corrections is affected by the choice of boundary conditions in the transverse direction. Specifically, the corrections are significantly smaller when periodic boundary conditions are applied in the transverse direction.  

This trend is particularly evident in the case of \( W = 3 \), as highlighted in Figure~\ref{fig:trimers_balanced}. The comparison clearly demonstrates the reduced influence of finite-length effects under periodic boundary conditions, underscoring the role of boundary configurations in shaping the observed trimer energy values.

Having established that trimer states do not exist in the quasi-one-dimensional regime of Hubbard ladders under the balanced case, we can now transition our focus to the imbalanced case. Without loss of generality, we assume \( t_{\uparrow} > t_{\downarrow} \), setting the stage for further exploration of the imbalanced trimer energy behavior and the associated physical implications. This expands the analysis to examine how hopping parameter imbalance affects the system's properties.

\begin{figure}[H]
\centering
\includegraphics[width=0.6\columnwidth]{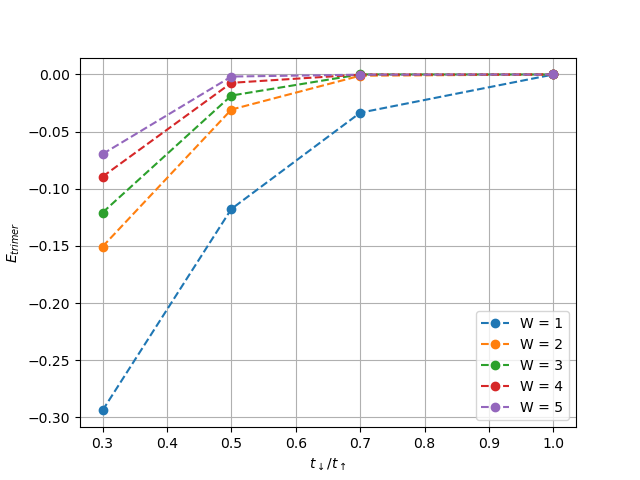}
\caption{Binding energy of a trimer \eqref{eq:trimer} for $W = 1,2,3,4,5$, $U = -7$ and open boundary conditions in transverse direction vs tunneling amplitude imbalance (mass imbalance) for different pseudospin projections. Filled circles are results of quadratic approximation (analogous to Fig.~\ref{fig:trimers_balanced}) of DMRG-computed trimer binding energies for $L$ up to $250$ to the limit $L \rightarrow \infty$ and dashed lines are guide to an eye 
\label{fig:trimer_energy_imbalanced}}
\end{figure}

From this point onward, we will focus on analyzing trimers composed of two spin-down electrons and one spin-up electron(two heavy, one light). Figure~\ref{fig:trimer_energy_imbalanced} presents the results, where the dots represent the quadratic approximation of the dependence \( E_{trimer}(1/L) \), calculated in a manner similar to the analysis shown in Figure 2. This approach provides a clear depiction of how the trimer energy evolves with the system's length.

It is easily seen that as the ratio of tunneling amplitudes decreases, the energy of the bound states, or trimers, increases. This indicates that a greater disparity between the tunneling amplitudes enhances the stability of the trimer states, as reflected in their higher energy values. Furthermore, the system's dimensionality plays a crucial role in determining the conditions under which bound states appear. The farther the system deviates from a strictly one-dimensional configuration, the greater the imbalance in tunneling amplitudes required to support the formation of bound trimer states. This observation highlights the interplay between dimensionality and tunneling asymmetry in shaping the energy landscape of trimers. It also underscores the sensitivity of the system's behavior to changes in both tunneling ratios and dimensional constraints.

We now turn our attention to the explicit measurement of superconducting correlation functions. 

In the one-dimensional case, correlations of this kind exhibit exponential decay when the filling fractions of fermions are commensurate, with heavy fermions being present in a 2:1 ratio. However, as soon as there is any deviation from commensurability in either direction, the trimer-induced suppression of superconductivity vanishes \cite{Orso2010trimers}. To ensure reproducibility, the parameters used in the calculations align with those in reference .\cite{Orso2010trimers}, specifically \( U = -5t_\uparrow \), with \( t_\uparrow = 1 \), \( t_\downarrow = 0.3t_\uparrow \), and \( L = 80 \). The ground-state wave function is computed for a fixed number of fermions with spins oriented up and down. This wave function is then utilized to construct the superconducting correlator in accordance with the form given in equation \eqref{eq:correlation}.

\begin{figure}[H]
\centering
\includegraphics[width=0.65\columnwidth]{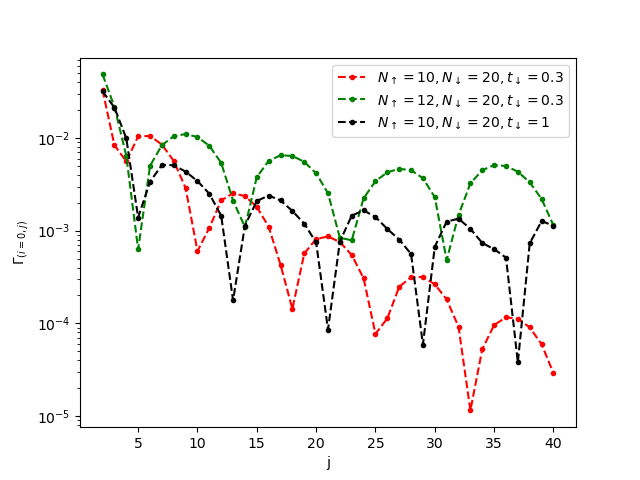}
\caption{Pair-pair correlation functions $\Gamma(i=0, j)$, in real space on a $(W=2)\times (L=80)$ ladder; 
$t_{\uparrow} = 1$. Correlations are measured along one of the legs and $i=0$ is the center of the leg. The scale on the $\Gamma$ axis is logarithmic.
\label{fig:pair-pair-corr-w2}}
\end{figure}

For the specific case of \( W = 1 \), our findings exhibit perfect agreement with the results presented in Ref.~\cite{Orso2010trimers}, providing robust validation of our approach. Figure~\ref{fig:pair-pair-corr-w2} illustrates the dependencies of the pair correlation function \(\Gamma(i=0, j)\) as a function of the distance \( j \) from the center of the leg, under three distinct configurations. The first configuration corresponds to commensurate carrier densities with \( t_{\downarrow} = 0.3 \), represented by the red line. The second configuration also involves commensurate densities but with a balanced tunneling amplitude, \( t_{\downarrow} = 1 \), represented by the black line. Finally, the third configuration considers incommensurate carrier densities with \( t_{\downarrow} = 0.3 \), represented by the green line. Analysis of the figure reveals that the pair correlations exhibit markedly different decay behaviors depending on the densities. For commensurate densities, the correlation values decay exponentially with distance, indicating a rapid suppression of long-range ordering. In contrast, for incommensurate densities or balanced tunneling amplitudes (\( t_{\downarrow} = t_{\uparrow} \)), the decay follows an algebraic trend, suggesting the persistence of correlations over longer distances. These results provide clear evidence that the presence of trimer states significantly suppresses long-range order in systems with commensurate carrier densities and this suppression is not restricted to strictly one-dimensional systems, but extends to quasi-one-dimensional geometries as well, highlighting the robustness of this phenomenon.

\begin{figure}[H]
\centering
\includegraphics[width=0.65\columnwidth]{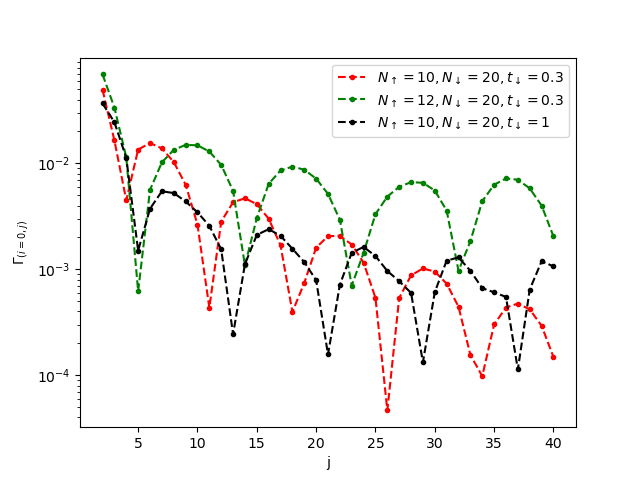}
\caption{Pair-pair correlation functions $\Gamma(i=0, j)$, in real space on a $(W=3)\times (L=80)$ ladder; 
$t_{\uparrow} = 1$. Correlations are measured along the central leg and $i=0$ is the center of the leg. The scale on the $\Gamma$ axis is logarithmic.
\label{fig:pair-pair-corr-w3}}
\end{figure}

For the \( W = 3 \) the interpretation of the results is significantly less straightforward compared to narrower systems. As shown in Figure~\ref{fig:pair-pair-corr-w3}, a noticeable trend can be observed regarding the rate of decrease of pair correlations. Specifically, for commensurate fermion densities, the decay of correlations is more rapid in the presence of unequal tunneling amplitudes (\( t_{\downarrow} \neq t_{\uparrow} \)) than in the case of incommensurate densities. This observation suggests that trimers play a role in suppressing long-range superconducting correlations under these conditions. However, when comparing correlation decay rates at the same carrier densities but with different \( t_{\downarrow}/t_{\uparrow} \) ratios, no definitive conclusion can be drawn about which tunneling ratio leads to a faster decline. This ambiguity implies that while trimers may exert some influence on the superconducting properties of the system, their effects are more nuanced and less pronounced in wider ladders.

To further investigate the influence of trimers and provide additional evidence supporting these conclusions, we extended our analysis to ladder lattices with four and five legs. By increasing the ladder width, we aimed to evaluate whether the observed trends persist and whether the trimer-induced suppression of long-range correlations becomes even less discernible as the system transitions from quasi-one-dimensional to more two-dimensional geometries. This approach ensures that our conclusions about the interplay between trimer states, carrier densities, and tunneling ratios are robust and applicable across systems with varying spatial configurations. These additional investigations are critical to assessing the limitations of trimer-induced phenomena in quasi-one-dimensional superconductivity and understanding how ladder geometry influences the strength and range of pair correlations.

\begin{figure}[H]
\centering
\includegraphics[width=0.8\columnwidth]{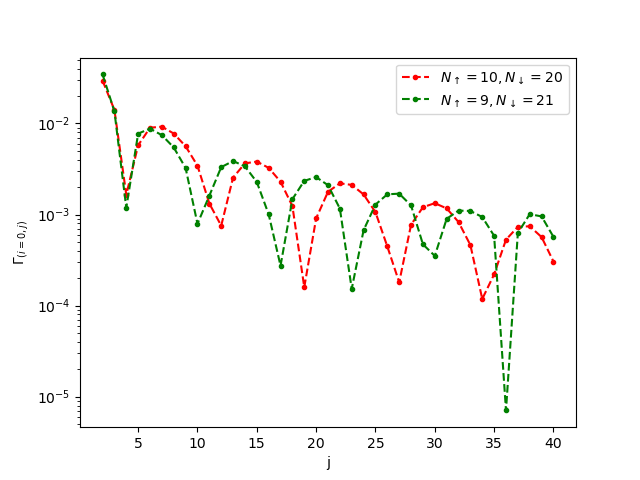}
\caption{Pair-pair correlation functions $\Gamma(i=0, j)$, in real space on a $(W=5)\times (L=80)$ ladder; 
$t_{\uparrow} = 1$ and $t_{\downarrow} = 0.3$,. Correlations are measured along the central leg and $i=0$ is the center of the leg. The scale on the $\Gamma$ axis is logarithmic.
\label{fig:pair-pair-corr-w5}}
\end{figure}

In Figure~\ref{fig:pair-pair-corr-w5} it is evident that the difference (if any) in the rate of decrease of correlations on a ladder with $W = 5, t_{\downarrow} = 0.3t_{\uparrow}$ for commensurate and incommensurate charge-carrier densities became almost unnoticeable, so we can conclude that with an increase in the number of legs (transition from 1D mode to 2D mode), the influence of trimers decreases.

\begin{figure}[H]
\centering
\includegraphics[width=0.8\columnwidth]{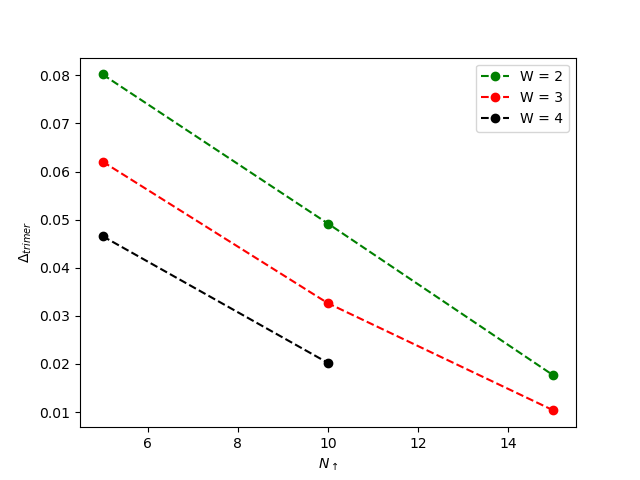}
\caption{Trimer gap for Hubbard ladders with open boundary conditions in transverse direction as a function of filling factor parametrized by $N_{\uparrow}$. $W=2,3,4$, $L=80$, $t_{\uparrow} = 1$, $t_{\downarrow} = 0.3$. 
\label{fig:trimer-gap}}
\end{figure}

In addition to the above-mentioned, it is worth noting that the influence of trimers decreases not only with the width of the ladder, but also with the filling fraction. For this, similarly to the work \cite{Orso2007}, one can consider the dependence of the trimer gap on the filling of the lattice at commensurate densities.

\section*{Summary}
We determined how the binding energy of trimers on Hubbard ladders depends on several factors, including ladder length, the number of legs (\( W = 2, 3, 4, 5 \)), the boundary conditions in the transverse direction (open or periodic), and the ratio of tunneling amplitudes for electrons with different spin orientations. For ladders with a width greater than two legs, this dependence has been analyzed for the first time. By extrapolating \( E_{\text{trimer}}(1/L) \), it was shown that, in the limit of an infinitely long ladder, the computed trimer binding energy is smaller than the approximation error (\(10^{-4}\)) for spin-independent tunneling amplitudes. However, it becomes significantly larger (around \(0.1\)) for \( t_\downarrow = 0.3t_\uparrow \). Additionally, it was found that finite-length lattice corrections are less pronounced when periodic boundary conditions are applied transversely.

The results suggest that the influence of three-particle bound states diminishes as the system transitions from a strictly one-dimensional configuration to a quasi-one-dimensional one. Furthermore, it was observed that as the filling factor increases, the trimer energy gap vanishes, indicating that the effect of trimers is only significant at low filling factors. 

\section*{Acknowledgments}
We are thankful to L. Shchur and Y. Budkov for fruitful discussions. We gratefully acknowledge support from the Basic Research Program of the HSE University, Russia. Numerical simulations were performed using HSE HPC resources~\cite{Kostenetskiy2021}.

\bibliographystyle{spmpsci}
\bibliography{references}

\end{document}